\documentclass{aa}  

\usepackage{graphicx}
\usepackage{physics,amsmath}
\usepackage{txfonts}
\usepackage{xcolor}
\usepackage{caption}
\usepackage{subcaption}
\usepackage{float}

\usepackage{natbib}
\bibpunct{(}{)}{;}{a}{}{,}
\usepackage[colorlinks=true, allcolors = blue]{hyperref}

\begin{document}

    \title{Spectral synthesis of 3D unified model atmospheres with winds for O stars}

    \author{L. Delbroek
          \and
          J.O. Sundqvist
          \and
          D. Debnath
          \and
          N. Moens
          \and
          F. Backs
          \and
          C. Van der Sijpt
          \and
          O. Verhamme
          \and
          P. Schillemans
          }

   \institute{Institute of Astronomy, KU Leuven, Celestijnenlaan 200D, bus 2401, 3001 Leuven, Belgium\\
             }

   \date{Received 15 July 2025/ Accepted  19 October 2025}
 
  \abstract  
   {Spectroscopic studies of massive and luminous O-type stellar atmospheres and winds have primarily been done by using one-dimensional (1D), spherically symmetric, and stationary models. However, both observations and modern theoretical models show that such O stars have highly structured and variable atmospheres and winds.} 
   {We present a first spectral synthesis based on three-dimensional (3D) time-dependent unified radiation-hydrodynamic (RHD) model atmospheres with winds for O stars.} 
   {We first carried out time-dependent, 
   3D simulations of unified O-star atmospheres with winds. We then used 3D radiative transfer to compute surface brightness maps for the optical continuum as well as integrated flux profiles for select diagnostic lines.
   To derive occupation numbers and source functions, an approximate non-local thermodynamic equilibrium method was used, as well as scattering source functions.} 
   {Our continuum intensity maps of a prototypical early O star ($\left<T_{\rm eff} \right> = $ 40.2 kK, $\log_{10} \left(\left<L_\star\right>/L_\odot\right) = $ 5.79) in the Galaxy reveal a highly variable and time-dependent surface, characterised by local emergent radiation temperature variations exceeding 10~000~K. Our averaged synthetic line profiles of optical absorption lines have large widths, 
   since the RHD simulations have large velocity dispersions
   in photospheric layers. Additionally, the absorption line equivalent widths are larger than for comparable 1D models. As such, to reproduce the 3D absorption line results in a corresponding 1D model, we need to apply isotropic Gaussian microturbulence, $\varv_{\rm mic} \sim 10-20 \, \rm km/s$, and macroturbulence, $\varv_{\rm mac} \sim 70 \, \rm km/s$, to the latter.
   First results using scattering source functions further demonstrate that characteristic features such as the softening of the blue edge of strong ultraviolet wind lines are qualitatively well reproduced by our models.}
   {Our 3D simulations clearly predict a highly structured and strongly variable O-star surface, in stark contrast with the smooth surfaces assumed by the 1D models currently used for quantitative spectroscopy of such stars. The first line profile results further suggest that several observed features are naturally reproduced by our models without the need to introduce ad hoc spectral fitting parameters. We also discuss how using 3D rather than 1D simulations as a basis for future studies may affect the derivation of fundamental stellar parameters such as surface gravities and chemical abundances.} 


   \maketitle
%

\section{Introduction}

  Traditionally, the assumptions of a spherically symmetric, stationary one-dimensional (1D) atmosphere and wind have been made in spectroscopic studies of O stars (for example \citealt{1D_model_source3, 1D_model_source1, 1D_model_source2, POWR3}). 
  However, theoretical studies have long indicated that the coupled envelopes, atmospheres, and line-driven winds (\citealt{Castor1975}) of O stars are highly structured and variable (due to convective and radiative envelope instabilities, as well as wind instabilities, \citealt{ Hearn, Stan_instabilities, Blaes_2003, Cantiello, Jiang_2015, Cassandra_2025, Key_2025}). 
  Other strong indications can be found in various observed phenomena; for example optical absorption lines that indicate very large `micro' and/or `macroturbulent' velocities (the latter even going up to > 100 km/s, see for example \citealt{Simon-Diaz2017}), optical emission lines that show line profile variability, and `wind clumping' \citep{macroturb_Conti, Eversberg_1998, Clumping_Puls, ML_2D, Simon-Diaz2017}. 
 
 Recently, we performed time-dependent, two-dimensional (2D) unified simulations of O-star atmospheres with winds using a flux-limiting radiation-hydrodynamical (RHD) finite volume modelling technique, accounting properly for the effects of line driving \citep{Dwaipayan}. In these simulations, opacities were computed using a hybrid approach \citep{Luka_2020} that combined tabulated Rosseland mean opacities with calculations of the enhanced line opacities expected for supersonic flows. The latter were here based on the \citet{Sobolev_1960} approximation. In these simulations then, structure formation can already be found just below the iron-opacity peak (located at approximately 200~kK, \citealt{Iglesias_1996}). By means of radiative acceleration exceeding gravity, local pockets of gas shoot up from these deep layers into the atmosphere above. Once in the upper atmosphere, these pockets then interact with the overlying line-driven wind outflow, creating a highly structured, turbulent atmosphere. This interaction also gives rise to large turbulent velocities in the photospheric layers of such unified atmosphere and wind simulations for O stars. The order of the turbulent velocities is $\sim$ 30-100~km/s, with higher values for models with higher luminosity-to-mass ratios. This is in general good agreement with observations of photospheric macroturbulence in O stars \citep{Simon-Diaz2017}, and also overall agrees well with results from recent independent three-dimensional (3D) massive-star modelling by \citet{Schultz_2023}, albeit without accounting for the effects of line driving and wind outflows. In that work, the authors used the Monte-Carlo radiative transport framework SEDONA \citep{SEDONA} to produce absorption lines from 3D RHD simulations (which were run using ATHENA++, \citealt{Jiang_2015, Schultz_2022}). 

In this paper, we present the first results from our 3D unified atmosphere and wind simulations of O stars. We focus here on the spectroscopic signatures of such simulations using 3D radiative transfer techniques to create synthetic images and spectra (building from the general code package presented in \citealt{Levin_2020, Levin}).

\section{3D unified atmosphere and wind simulations of O-type stars}
\subsection{RHD simulation method} \label{RHO_sim_method}

Previously, \citet{Dwaipayan} studied unified atmosphere and wind simulations of O-type stars using the RHD module from \citet{nico_2022a} of MPI-AMRVAC \citep{Xia_2018, Keppens_2023} in 2D. We took their prototypical `O4'  model, and used the same numerical set-up to calculate new simulations by solving the time-dependent RHD equations on a finite volume
in three spatial dimensions, including correction terms for spherical divergence (see \citealt{nico_2022a}). 
This means that we solved the equations of mass, momentum, and energy conservation, including the effects of the radiation field in both the total momentum (radiation force) and energy (radiative heating and cooling). Gravity was included for a fixed point stellar mass, $M_\star$. The radiation field was computed from the frequency-integrated time-dependent radiation energy equation in the co-moving frame. The necessary closure relations between radiation energy density, pressure, and flux were obtained analytically with a flux-limiting procedure that preserves the optically thick and thin limits. Mean opacities include Rosseland mean opacities and line driving effects in the same way as was first outlined in \citet{luka_2022}, and as in previous work we assumed that the flux, energy, and Planck mean opacities were the same. For further details, including exact equations as well as specifications of initial and boundary conditions, see \citet{nico_2022b, nico_2022a, Dwaipayan}. The radial direction extends (approximately) thrice the lower boundary radius, $R_0$. This $R_0$ is defined as the radius in which the temperature in the initial conditions reaches 500~kK, here $\simeq 13.54~ R_\odot$. Both the tangential directions cover 0.2~$R_0$. Radially, the box has 128 grid points for the lowest resolution, and both tangential directions have 16 grid points. The highest level of refinement is level four, which has a resolution that is a factor of $2^3$ higher than the lowest resolution (first level).

\subsection{Basic characteristics of early O-star 3D model}

 \begin{figure}[h!]
   \centering
   \includegraphics[width=0.95\hsize]{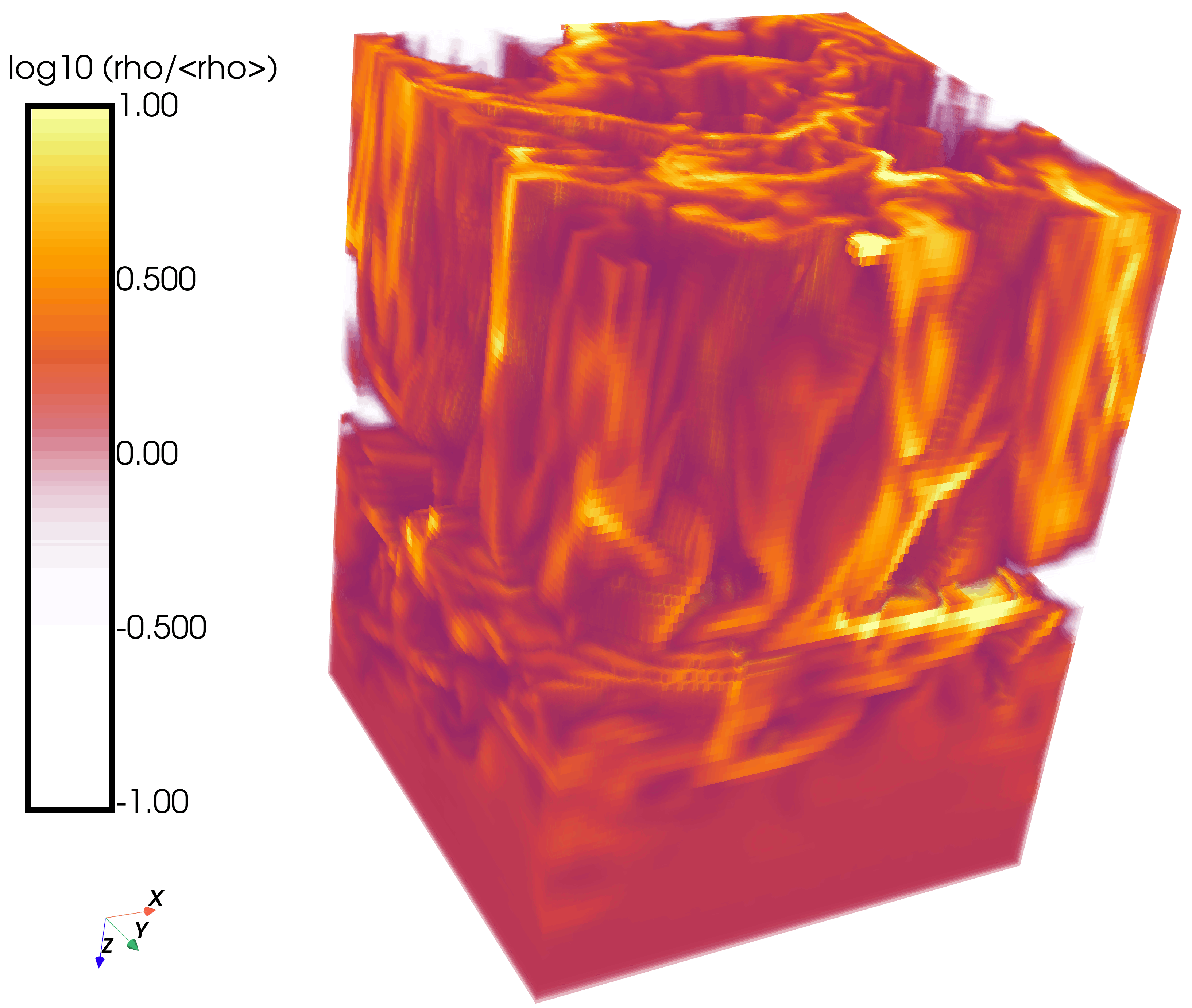}
      \caption{Volume rendering of relative density (displayed colours are $\log_{10} \rho/\langle \rho \rangle$) for the 3D model atmosphere and wind. Angle brackets denote lateral averaging at each radial position.}
         \label{Fig:rho_volume}
   \end{figure}

\begin{figure*} 
       \centering
\includegraphics[width=\linewidth]{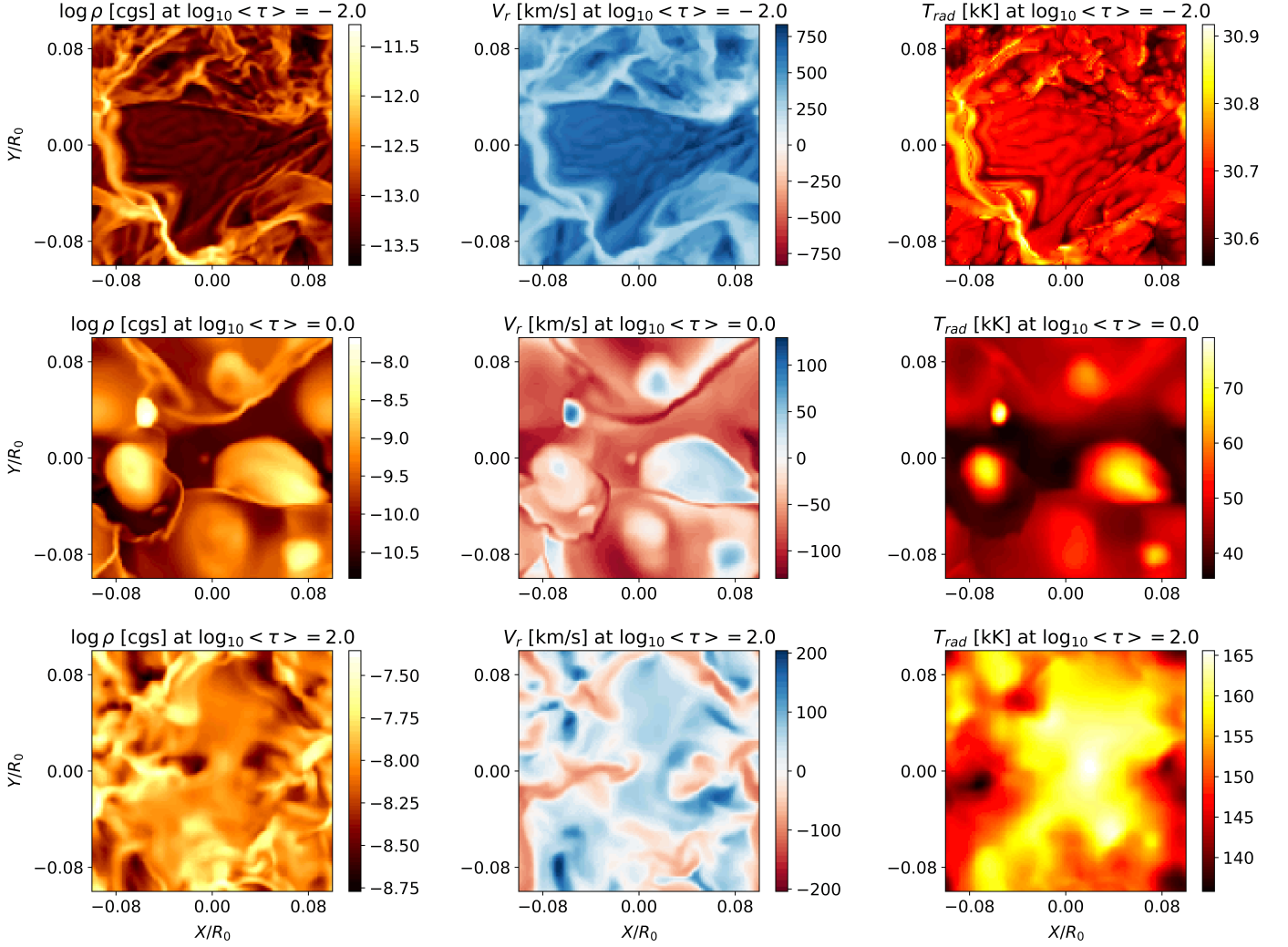}
      \caption{Maps of logarithm of density, radial velocity, and radiation temperature, in three selected vertical planes with laterally averaged continuum optical depth according to the legends. The lateral X and Y axes are labelled in units of $R_0$. 
      }
         \label{3d_slice} 
   \end{figure*}

As in the 2D simulations by \citet{Dwaipayan}, structure starts appearing in our 3D model just below the iron-opacity peak, where the gas and radiation temperature are $\sim$ 200~kK. 
As energy transport by convection in these atmospheres is inefficient \citep{Dwaipayan}, the very sensitive dependence of the Rosseland mean opacity on temperature around the iron opacity peak yields local pockets of gas that are radiatively accelerated exceeding gravity ($\vb*{g}_{\rm rad} > \vb*{g}_{\rm grav}$). As a result, these pockets shoot up with velocities higher than the gas sound speed (however, less than the gas+radiation sound speed, see \citealt{Dwaipayan}) into the upper and cooler atmosphere. 
There, Rosseland mean opacities are lower than in the deeper layers, and the gas is decelerated. While some gas parcels fall back into the deep atmosphere, others become subject to line driving and re-accelerate to launch a supersonic wind outflow. This complex interplay creates a very turbulent atmosphere characterised by large density and temperature fluctuations and high velocity dispersions. A more complete analysis of this process has been explored in \citet{Dwaipayan} for a set of 2D O-star models, and \citet{nico_2022b} for a 3D stripped, hot dwarf model.

Fig. \ref{Fig:rho_volume} shows a volume rendering of relative density in a selected part of the 3D atmosphere, taken at a snapshot well after the simulation has adjusted from the initial conditions. Relative density here means that the displayed colours are $\log_{10} \rho/\langle \rho \rangle$ where angle brackets denote lateral averaging at each radial position. The 3D rendering shows the part of the atmosphere from just below the `iron bump' where inhomogeneities first occur, across the turbulent and variable photosphere, and into the outflowing line-driven wind (up to 1.5 above the average stellar surface). The figure clearly illustrates the turbulent, filamented, and variable nature of the O-star envelope and atmosphere.
This can also been seen in Fig. \ref{3d_slice}, in which we see one snapshot sliced at different laterally averaged electron scattering optical depths (and hence different heights above the inner boundary of the snapshot). The upper three panels correspond to a wind part of the snapshot, the middle three panels are around the optical photosphere and the bottom three deep down in the atmosphere. Inspecting the panels in Fig. \ref{3d_slice}, it can be seen that these three different regions have different structures, velocities, and temperature fluctuations. As noted above, in the deep atmosphere around the iron bump, opacity has a very strong temperature dependence. On the other hand, Rosseland mean opacities are also positively correlated with density, meaning that we sometimes observe blobs of high-density gas around the photosphere that are radiatively accelerated (see also \citealt{Jiang_2018}) and thus have rather high positive velocities (middle panels). By contrast, radiative acceleration due to line driving is negatively correlated with density, implying that high density wind regions (upper panels) do not experience efficient line driving but are rather dragged along (with relatively low velocities) by the surrounding rapidly accelerating low-density wind (see also discussions and figures in \citealt{nico_2022b, Dwaipayan}). Moreover, radiation temperature fluctuations are highly damped in the wind parts of the simulation as compared to the large fluctuations around the photosphere and in the deeper layers. 

Table \ref{table:Models} lists fundamental average parameters of the 3D RHD model. Lateral averages are here taken over 65 snapshots each separated in time by $\sim 500$ seconds, taken well after the simulation has relaxed from initial conditions. Photospheric parameters are then computed at the average vertical position in the atmosphere corresponding to continuum $\langle \tau_c \rangle = 2/3$, whereas the average mass-loss rate, $\langle \dot{M} \rangle$, and maximum wind speed, $\langle \varv_{\rm max} \rangle$, are taken close to the outer boundary in the outflowing parts of the simulation.\footnote{We note that since the average wind is still somewhat accelerating at our outermost grid point (see also \citealt{Dwaipayan}) we need to distinguish between this maximum average wind speed in the simulation and a final terminal wind speed 
$\varv_\infty \equiv \varv_r(r \rightarrow \infty)$.} Finally the Eddington luminosity is defined as usual, $L_{\rm Edd} \equiv 4 \pi G M_\star c/\kappa_e$, where $\kappa_e$ is the electron scattering opacity, which we also take to be the sole continuum opacity in the computation of the optical depth. Similar to \citet{Dwaipayan}, we computed that the characteristic dynamical timescale in the deeper atmosphere is of the order of $t_{d,a} \sim$ 1000~s, while the dynamical timescale in the wind is of the order of $t_{\rm d,w} \sim$ 10~000~s.
    \begin{figure*} 
           \centering
    \includegraphics[width=\linewidth]{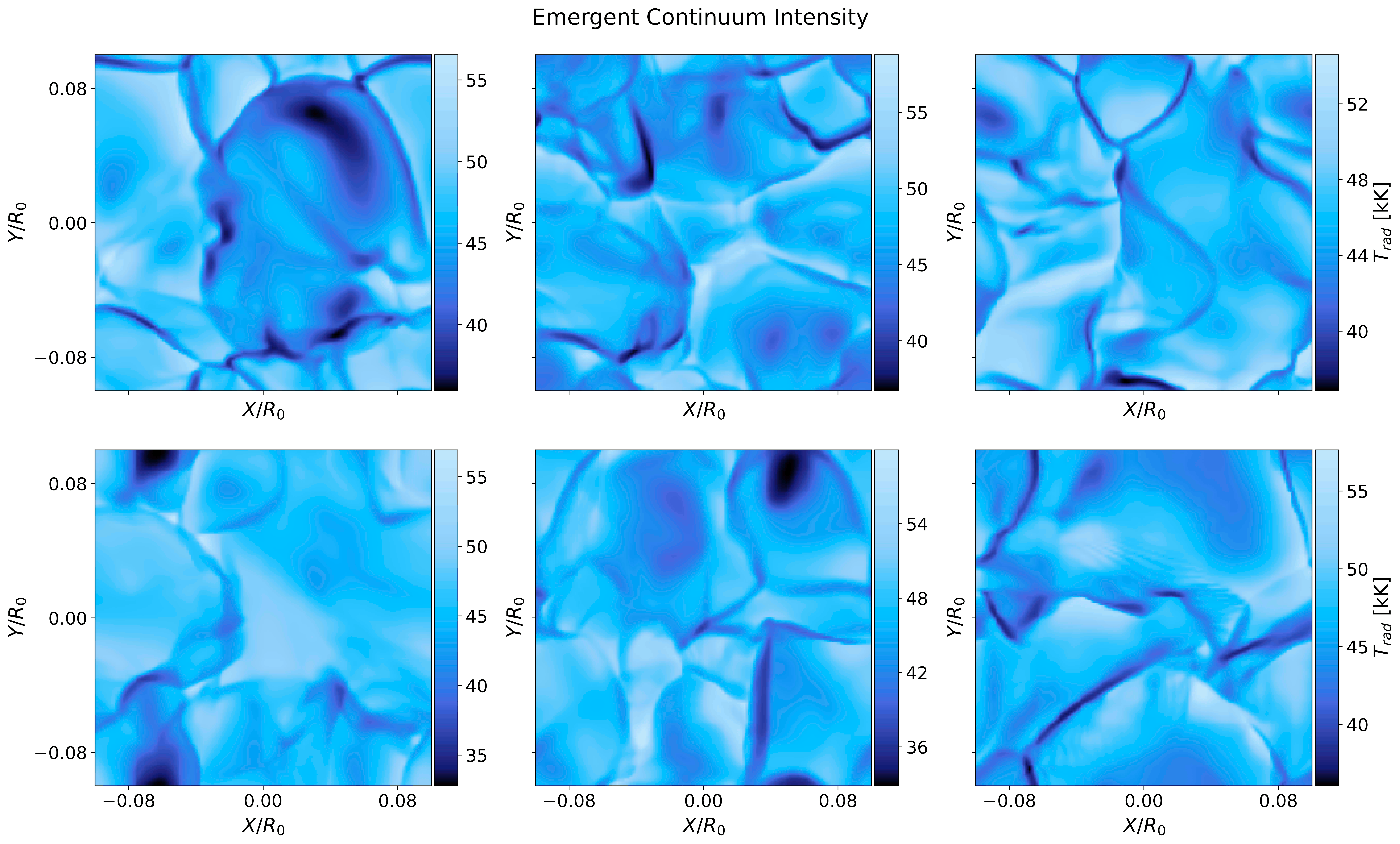}
          \caption{Emergent optical continuum intensities of local patches on the 3D O star model, for six different snapshots. Each panel here corresponds to one $\mu$=1 (local surface normal and line of sight are aligned) 3D patch simulation, and is showing the complete patch. Colour bars display local emergent radiation temperature defined through $I_{\nu} \equiv B_\nu(T_{\rm rad})$. The spatial extent of each X and Y axis is $0.2~R_0$.}
             \label{Fig:local_surf} 
    \end{figure*}
    
\begin{table*}
\caption{Fundamental parameters for the $\langle 3\rm{D} \rangle$ O-star model studied in this paper. 
}       
\label{table:Models}      
\centering          
\begin{tabular}{c | c c c c c c c c }  
\hline\hline       
Model & $\left<T_{\rm eff} \right> \rm [kK]$ & $M_\star/M_\odot$ & $ \langle R_\star \rangle/R_\odot$ & $\log_{10} \left(\left<L_\star\right>/L_\odot\right)$ & $\left<L_\star\right>/L_{\rm edd}$& $ \log_{10} \left<g_\star\right>$ &  $\log_{10} \left<\Dot{M}\right>  \ [M_\odot/yr] $ & $\left< \varv_{\rm max} \right> \rm [km/s]$ \\ 
\hline                    
   $\rm{O}4$ & 40.2  & 58.3 & 16.2 & 5.79 & 0.27 & 3.79 & -5.90 & 1800 \\
\hline                  
\end{tabular}
\tablefoot{From left to right, the columns display model name, effective temperature, stellar mass, radius, luminosity, Eddington ratio, surface gravity, mass-loss rates, and maximum velocity. Angle brackets denote averaged quantities as explained in the text.}
\end{table*}

\section{Methodology of radiative transfer calculations} \label{methodology} 
    Three-dimensional radiative transfer techniques (building from \citealt{Levin_2020}) 
    are now applied to the 3D unified atmosphere and wind simulations of O stars described above, more specifically on the whole domain of the spherical models, constructed from the 3D unified atmosphere and wind simulations. The construction of these spherical models is outlined in section \ref{3D_building}.
       
    \subsection{Equations of radiative transfer}
    Specific intensities were obtained by solving the time-independent equation of radiative transfer:  
    
     \begin{equation}
        \textbf{n} \nabla I_\nu = \eta_\nu - \chi_\nu I_\nu = \chi_\nu (S_\nu - I_\nu),
         \label{Time_independent equation of radiative transfer}
    \end{equation}

     where $I_\nu $ is the observer's frame specific intensity at frequency $\nu$, $\eta_\nu $ the emissivity, $\chi_\nu$ the opacity, $S_\nu $ the source function, and \textbf{n} the observer’s direction. 
  In the main part of this work we assumed that opacities and source functions could be calculated locally, using an approximate non-local thermodynamic equilibrium (aNLTE) technique following \citet{Lucy_Abbott_1993, Springmann_1997, Puls_2000}.  
  This method approximately corrects level population numbers computed in local thermodynamic equilibrium (LTE) by using potentially different radiation and gas temperatures, as well as a modified spherical dilution factor. 
  This modified spherical dilution factor is computed from the mean photospheric radius and optical depth, as is explained in, for example, \citet{Springmann_1997}. Since the aim of the code is to work with non-monotonic velocity fields (which are fully accounted for, see  \citealt{Levin_2020}), the equation of radiative transfer is solved in the observer's frame. This is done because a frequency-dependent comoving-frame formulation is very complicated to implement when dealing with highly non-monotonic velocity fields.
  Local radiation and gas temperatures were directly taken from the RHD simulation described in the previous section, and the dilution factor was computed from the radially averaged optical depth scale of the model. For continuum opacities and source functions, we assumed a constant Thomson scattering opacity and a Planck function using the local gas temperature. In Sect. \ref{First_line_profile_results_including_scattering}, however, we also provide first test results from computation of scattering-dominated source functions. Boundary conditions are as follows \citep{Levin_2018, Levin_2020}: if rays originated from the stellar core the radiation at the lower boundary was set to $I_\nu^+=B_v(T_{rad},\nu)$, with the other quantities obtained from trilinear interpolation from the 3D model atmosphere. Inside the core, we used $I_\nu^+ = S_L = S_C$ (with $S_L$ and $S_C$ the line and continuum source functions, respectively).  
  At the outer boundary, the intensities that come from outside this boundary were set to zero and the intensities coming from inside were calculated with our RT scheme. For rays that do not intersect the core, these boundary conditions were applied at both sides of the hemisphere. Finally, in this work we only treated single lines and we further assumed pure Doppler profiles with the widths set by the local thermal speed of the considered ion.

\subsection{Emergent intensities and fluxes}
     
The specific intensity found from equation (\ref{Time_independent equation of radiative transfer}) could be used to determine surface intensities of the 3D models. 
This emergent intensity was then integrated over the projected surface to obtain emergent fluxes. For planar models this is straightforward, but some care must be taken when accounting for sphericity effects. We used a cylindrical co-ordinate system (for example, \citealt{Sundqvist_2012}, \citealt{Levin}):

     \begin{equation}
 F_\nu = \frac{1}{d^2}  \int_{0}^{2\pi} \int_{-R_{\rm max}}^{R_{\rm max}} I_{\nu,n}(p, \zeta, z = R_{\rm max}) \,p\, \mathrm{d}p\,\mathrm{d}\zeta  ,
  \label{Emergent flux profiles}
   \end{equation}
  
    with $d$ being the distance from the observer, $R_{\rm max}$ the maximum size of the emitting object, $(p,\zeta)$ the cylindrical co-ordinates describing the projected disc of the emitting object perpendicular to the observer’s direction, \textbf{n}, and $I_{\nu,n}(p, \zeta, z = R_{\rm max})$ the emergent specific intensity into that direction evaluated at a distance $z = R_{\max}$. The z axis is aligned with the direction to the observer. The geometry is illustrated in Fig. 1 of \citet{Levin}. 
    Radiation is presumed to be free-streaming outside of the simulation domain, and as such the intensities at the outer radius do not change until the projected surface at $z=R_{\rm max}$. If the rays go through the core, our limits are not $-R_{\rm max}$ to $R_{\rm max}$, but the core radius up to $R_{\rm max}$.

    \subsection{Building 3D spherical models} \label{3D_building}

\begin{figure}[h!]
   \centering
   \includegraphics[width=0.95\hsize]{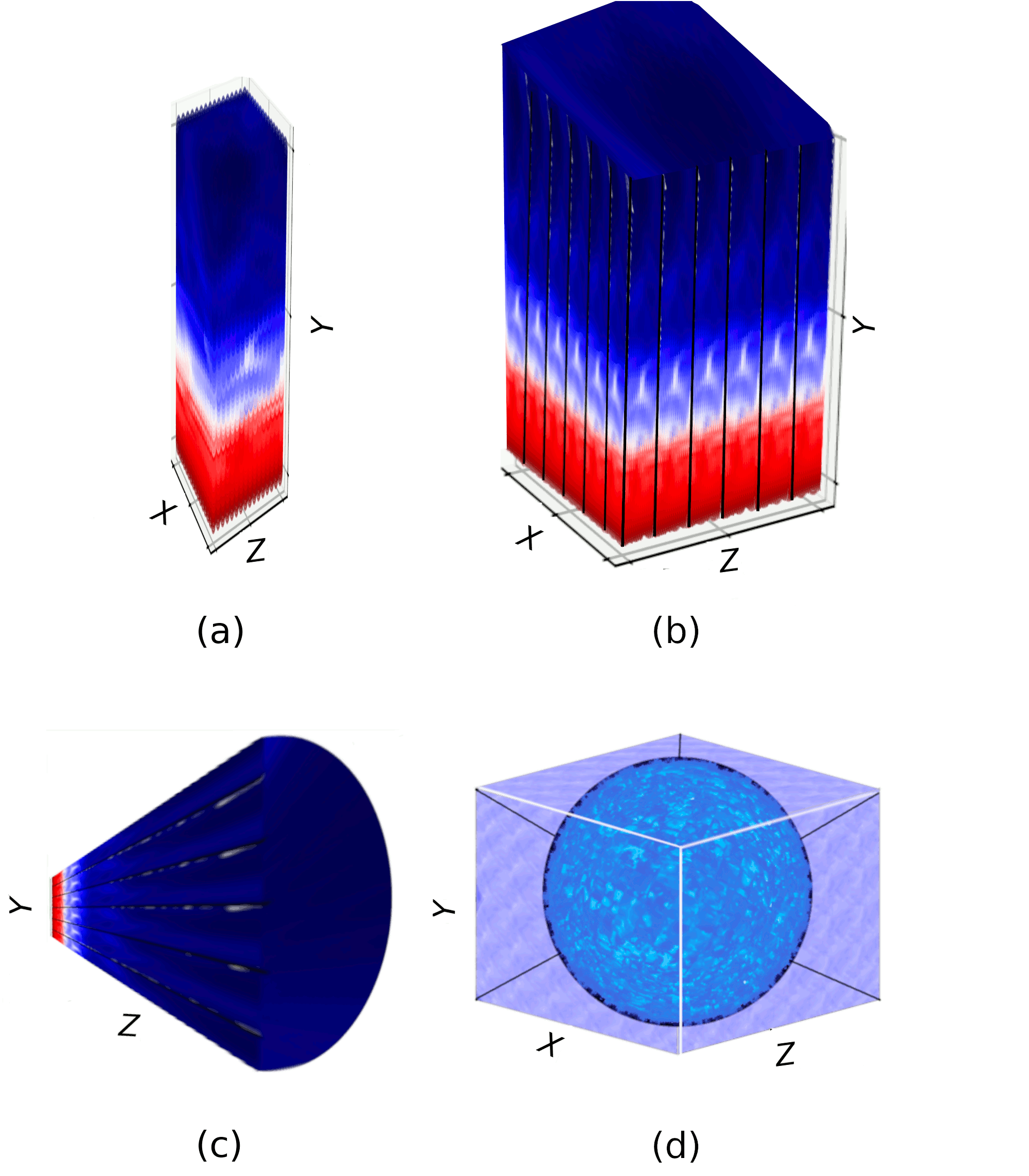}
      \caption{Figurative illustrations of (a) the snapshots (3D unified atmosphere
and wind simulations of O-type stars), (b) the stacked snapshots (stacked snapshot array), (c) one of the six sectors that is constructed from our stacked snapshot array and (d) the sphere that is constructed. For this illustration only one snapshot was used; hence a repetitive pattern is visible. The colour represents an arbitrary field in the simulation. }
         \label{Vis_all4_in_one}
   \end{figure}

    Before computing emergent flux line profiles, we reconstructed a global 3D spherical model from our local box-in-star RHD simulations.
    Since the width of the RHD simulations is much smaller than the radius of the star (as can be seen in Figures \ref{3d_slice} and \ref{Fig:local_surf}), we need many (in this specific case 100) simulation boxes (from here on called snapshots) to cover the full stellar and wind surface. Figure \ref{Vis_all4_in_one}a gives a figurative illustration of how typical snapshots look (shape-wise). Once the snapshots were chosen from which the 3D spherical model would be constructed, the selected snapshots were stacked together (see Fig. \ref{Vis_all4_in_one}b for an illustration of this process) and transformed into a sphere using cube-sphere transformations.
    The cubed-sphere co-ordinates split the sphere into six separate sectors (see Fig. \ref{Vis_all4_in_one}c), where each of the sectors has its own local co-ordinate system $(r,\xi,\zeta)$. Here, $r$ is the radial co-ordinate and $\xi$ and $\zeta$ are two angular co-ordinates. The transformation from our pseudo-planar co-ordinates from the RHD simulations to the cubed-sphere co-ordinates, happens by interpreting the radial co-ordinate from the simulations as the radial co-ordinate on the cubed sphere, and interpreting the two lateral co-ordinates as the two angular ones.  
    A final transformation turned this into a spherical star as illustrated in Fig. \ref{Vis_all4_in_one}d. 
    See for example \citet{RONCHI199693, Cube_sphere_trans} for more details on the cube-sphere transformation process. 
    
     A justified concern that should be noted is that the cube-sphere transformation (going from Fig. \ref{Vis_all4_in_one}b to Fig. \ref{Vis_all4_in_one}d) stretches and distorts the atmospheric structures. This distortion grows linearly with the radius as we interpret the lateral co-ordinate in the RHD simulations as an angular co-ordinate in the construction of the sphere. Most importantly, however, the radial distance from the stellar core for each point stays the same. Sphericity effects have been accounted for as much as possible already in the RHD simulations (see for example \citealt{nico_2022a}, Appendix A) and as such the radial structure is consistent.

            \begin{figure}[h!]
   \centering
   \includegraphics[width=0.95\hsize]{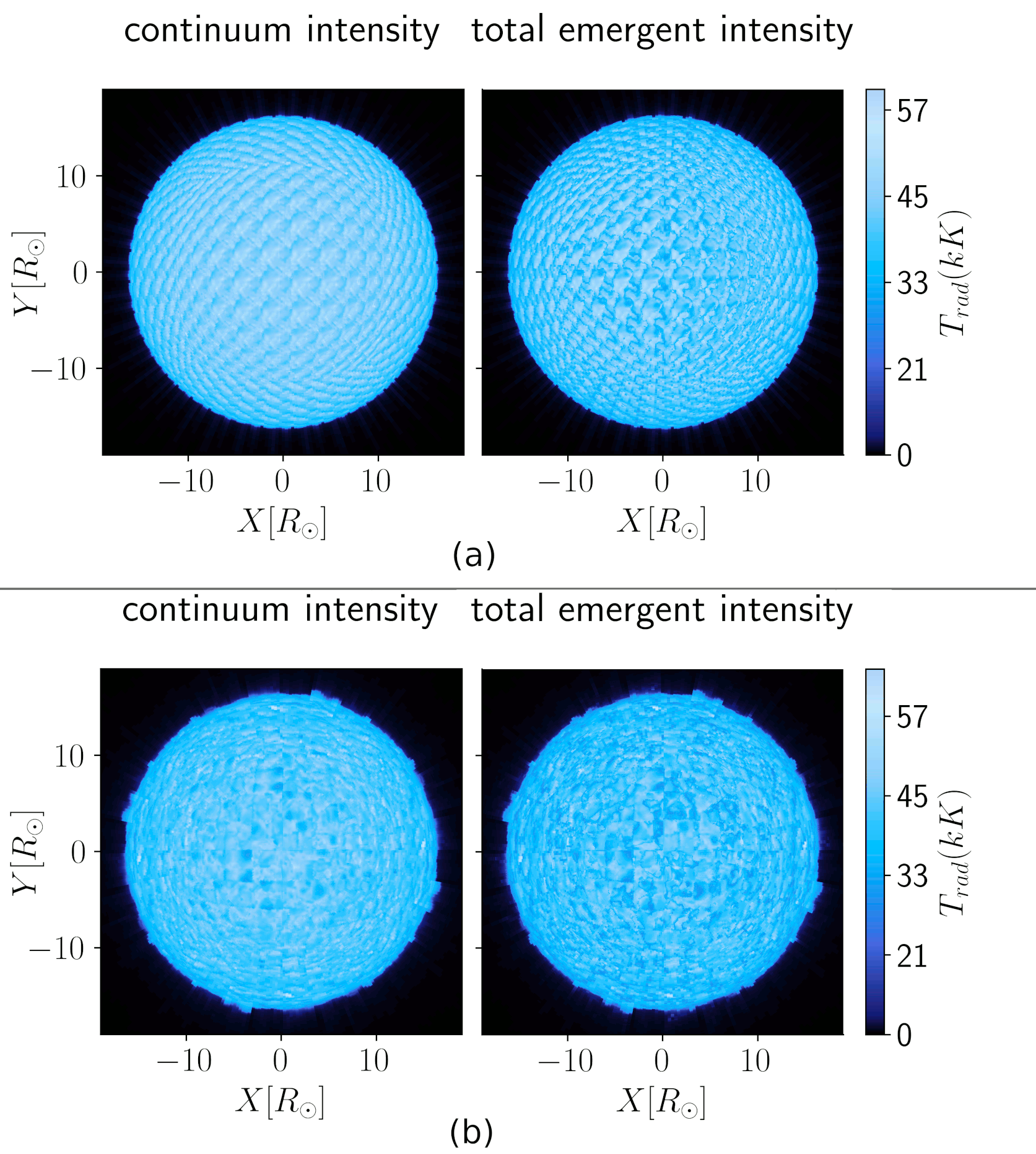}
      \caption{Surface brightness plots of the O-star simulations described above, at the wavelength of the O III 5594 line: (a) for a snapshot-sphere model, (b) for a mixed-sphere model.
              }
         \label{Surf_bright_O_III_snapshot_sphere_mixed_sphere_illustration}
   \end{figure}

    In this work, we used a set of 65 snapshots, each spaced approximately $0.5~ t_{\rm d,w}$ in time. 
    We built two kinds of 3D spherical models. First, we created `snapshot-spheres'. For these models, one snapshot was copied 100 times to construct our stacked snapshot array, and in total 600 times to create our full 3D spherical model (since the stacked snapshot array is copied six times to form our sphere). Second, we created `mixed-spheres'. For these models, we mixed our 65 snapshots to build our spherical model. We created 20 mixed-sphere models (in which we made sure that the snapshot placement was different from the other mixed-sphere models). Note that since we needed 100 snapshots to assemble our stacked snapshot array, we allowed that snapshots be used (at most) twice during this construction process. An important detail to keep in mind here, is that both types of models (mixed- and snapshot-spheres) used the exact same set of snapshots during their formation process; that is, the snapshot-spheres and mixed-spheres used the same time series. During the spherical model construction and the spectral synthesis process, we degraded the 3D information, and also transformed it onto other grids. The final cylindrical grid on which the formal integral was solved, has 397, 157, and 800 points in $(p, \zeta, z)$, respectively. We tested that doubling this resolution has a negligible impact on the final resulting flux profiles.
    
\section{The resolved O-star surface}

Before investigating the O-star surface of our spherical constructed models we first examine the individual snapshots. These are not subject to the deformations and stretching discussed in section \ref{3D_building}.

    \subsection{Continuum variation of local patch on O-star surface} \label{patch_surface}

           \begin{figure*}
       \centering
            \includegraphics[width=\textwidth]{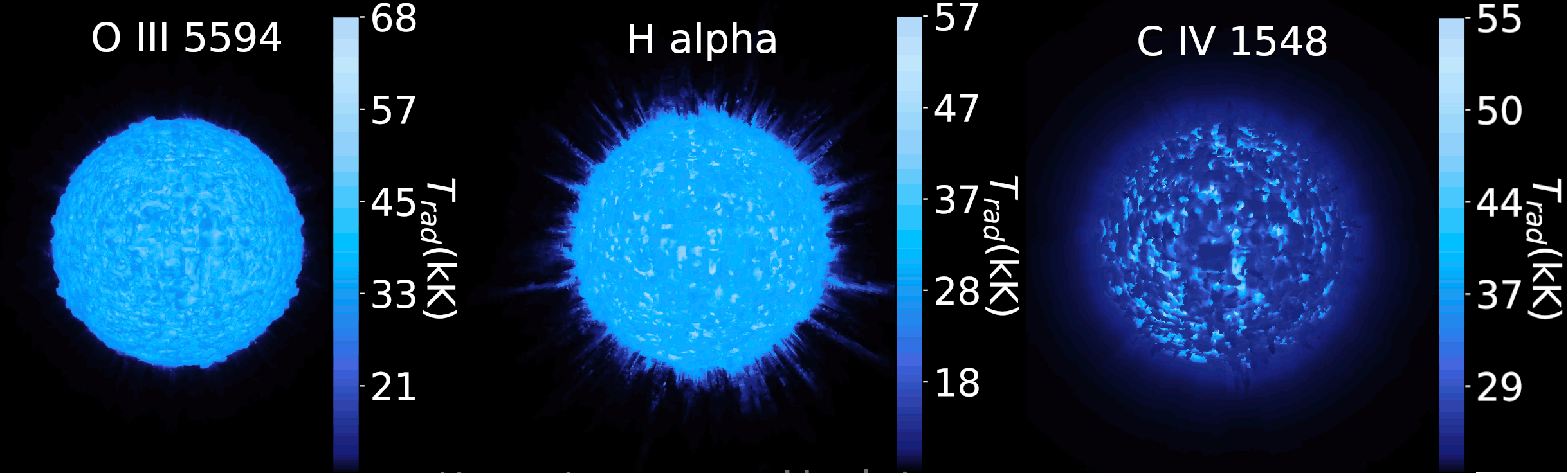}
      \caption{Surface brightness illustrations showing the emergent specific intensity as seen by an observer in a certain direction for the O III 5594 $\AA$ line, the H alpha line, and the C IV 1548 $\AA$ line at the line centre.}
         \label{Pretty_surf_bright}
   \end{figure*}

          \begin{figure}[h!]
   \centering
   \includegraphics[width=0.95\hsize]{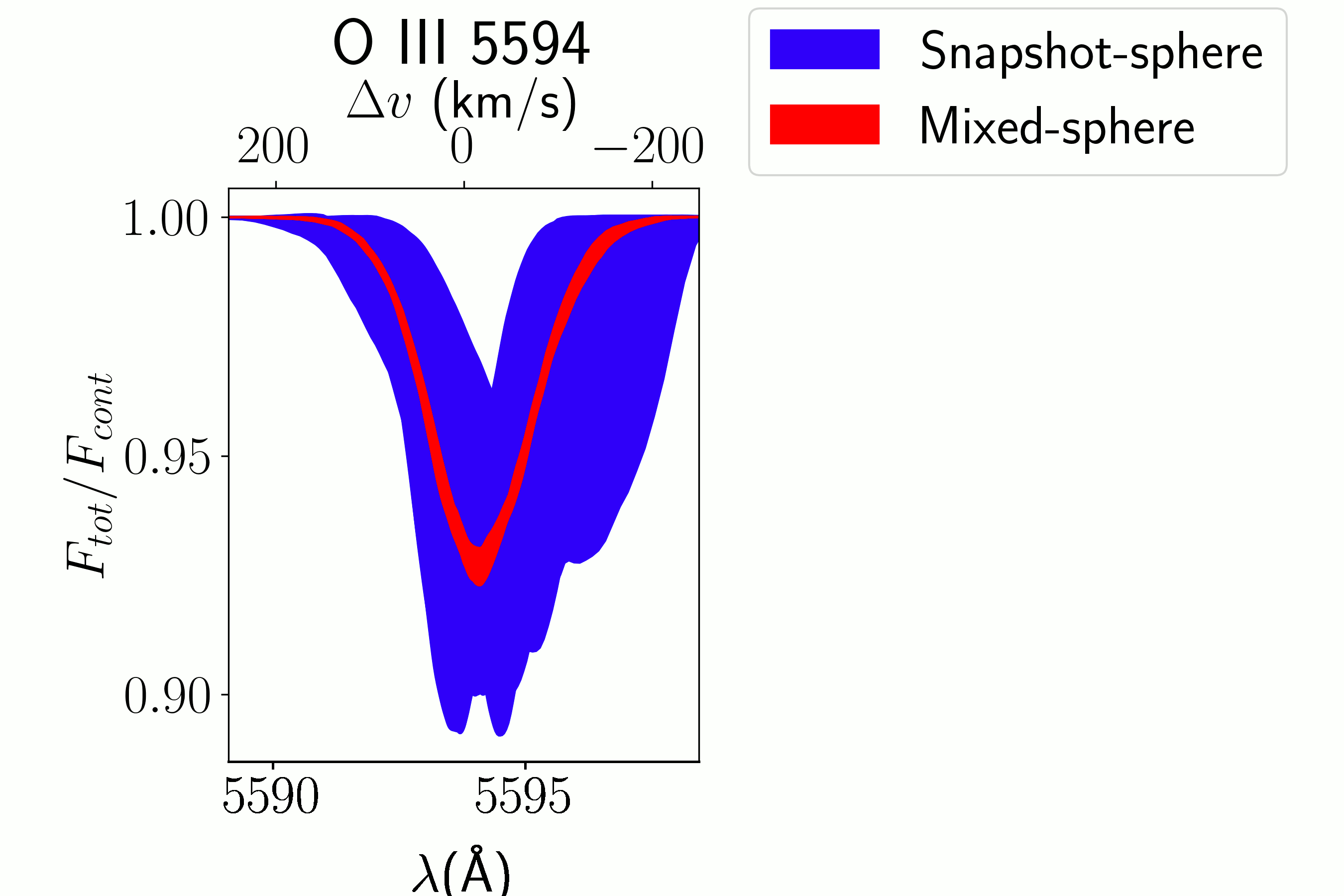}
      \caption{Coloured-in regions showing variability for the O III 5594 $\AA$ line. The blue corresponds to the variability of the snapshot-sphere models (65 models in total), while the red corresponds to the variability of the mixed-sphere models (20 models in total).
              }
         \label{colored_in_O_III_snapshot_sphere}
   \end{figure}
   
    Figure \ref{Fig:local_surf} displays emergent intensities for six snapshots of our 3D simulation, computed for a typical continuum wavelength in the optical assuming constant electron scattering opacity. The snapshots are separated by $\sim$ 2000 seconds (which is in between the $t_{\rm d,a} \sim$ 1000~s and $t_{\rm d,w} \sim$ 10~000~s) and taken well after the model has relaxed and adjusted itself from the initial conditions. The figure shows how the O-star surface is characterised by very large spatial and time fluctuations; labelled with the intensities expressed as a radiation temperature of $I_{\nu} \equiv B_\nu(T_{\rm rad})$, typical surface variations are as high as of the order of $\sim 10^4$~K. Since the total spatial dimension of these local patches is 0.2~$R_0$, the sharp lanes as well as the larger bubble-like features visible in the figure are small in comparison to the star itself. Overall, this reveals a dynamically very active surface in stark contrast with the homogeneous surface assumed by present-day standard 1D models used for spectroscopic studies of O stars.   
    
    \subsection{Surface brightness maps from spherical models}

    Using the 3D spherical surface volume reconstruction method described in Sect. \ref{3D_building}, Figs. \ref{Surf_bright_O_III_snapshot_sphere_mixed_sphere_illustration}a and \ref{Surf_bright_O_III_snapshot_sphere_mixed_sphere_illustration}b next show two examples of simulated surface brightness plots for the complete star. 
    The first panel in each figure corresponds to the continuum intensity, the second panel to the continuum + line total emergent intensity. 
    We note the visual difference between Figs. \ref{Surf_bright_O_III_snapshot_sphere_mixed_sphere_illustration}a and \ref{Surf_bright_O_III_snapshot_sphere_mixed_sphere_illustration}b; the former displays a repetitive pattern, as the same snapshot was copied over the entire sphere (the `snapshot-sphere' models). The latter, on the other hand, filled in the sphere using 65 distinct snapshots and as such avoids the regular patterns. 
    Because of this, we focus primarily on these `mixed-sphere' models in the rest of this work. For example, Fig. \ref{Pretty_surf_bright} displays the emergent intensity at line-centre across the surface for the three key diagnostic lines O III 5594 $\AA$, H$\alpha$, and C IV 1548 $\AA$. These lines were chosen for Fig. \ref{Pretty_surf_bright} because they represent three different kinds of important diagnostics in O stars (photospheric absorption line, wind-influenced recombination line, UV resonance wind line). As we in practice cannot resolve the O star surface in these wavelengths, the images in Fig. \ref{Pretty_surf_bright} are primarily for illustration purposes. Moreover, since these lines were calculated using an aNLTE approach, particularly the C IV 1548 $\AA$ line has to be treated with care as this line most definitely has a very scattering-dominated source function (see Sect. \ref{First_line_profile_results_including_scattering}). Nonetheless, what already can be clearly seen from these images is that this (mixed-sphere) star is very non-uniform, which is in contrast with the typical assumption that massive stars are pretty smooth and stable at the photosphere.

  \begin{figure*} 
       \centering
\includegraphics[width=0.84\linewidth]{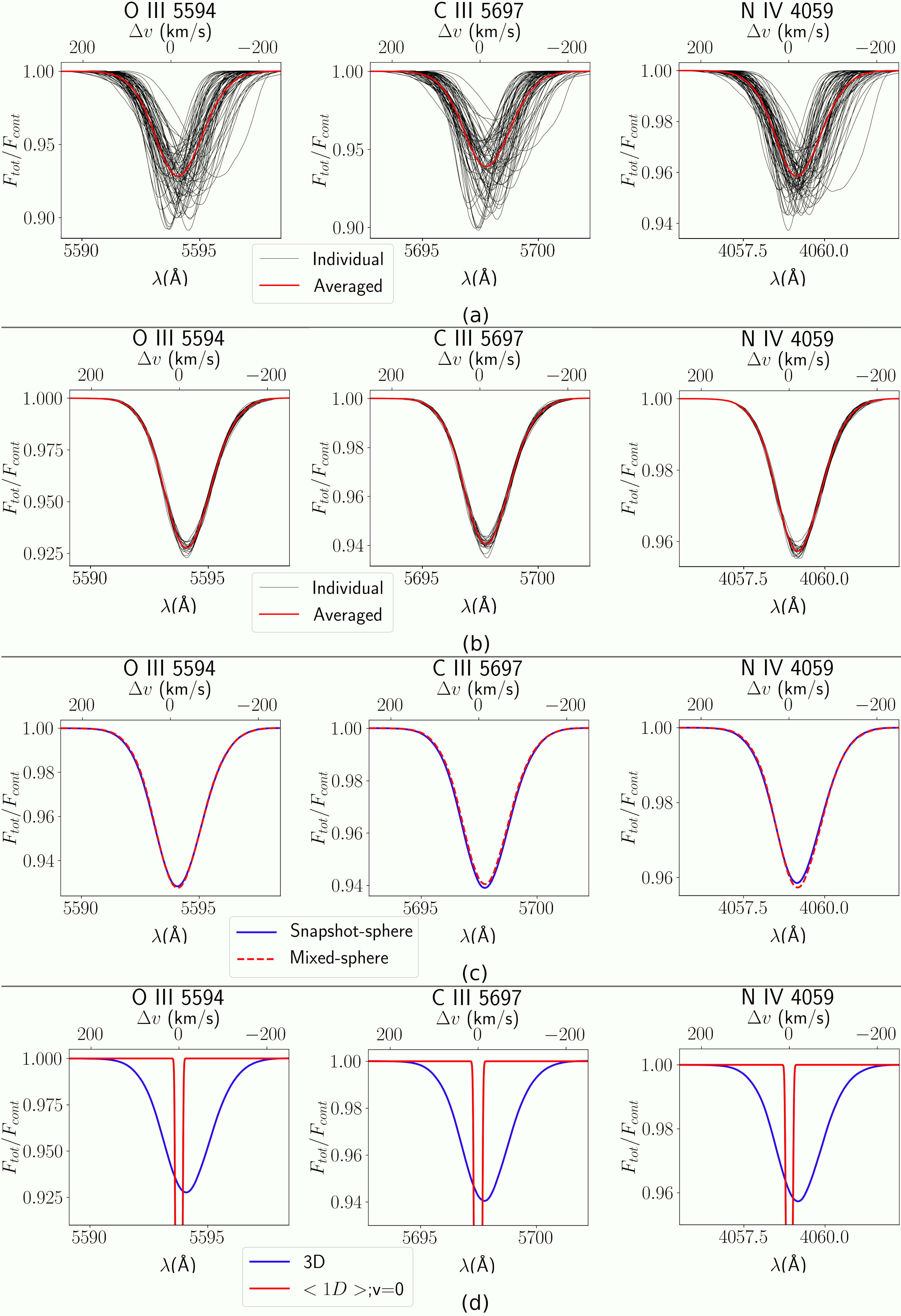}
      \caption{O III 5594 $\AA$ line, the C III 5697 $\AA$ line and the N IV 4059 $\AA$ line for the 3D O-star simulation, (a) using 65 snapshot-sphere models (the red lines show the result when all lines are averaged), (b) using mixed-sphere models (the red lines show the result when all lines are averaged), (c) showing averaged lines from both the snapshot-spheres (blue line) and mixed spheres (red line) and (d) showing averaged lines for the mixed-sphere models for both the `3D' (unaltered) models and the 1D averaged, velocity-field-turned-off models.}
         \label{BIG_FIG} 
   \end{figure*}

\section{Line profile results and trends}

\subsection{Line profile variability}
Spectroscopic line profile variability has been observed in O-type stars (for example \citealt{VAR3, VAR2, VAR1}). In our simulations, variability occurs naturally. 
Hence, let us first find approximate upper and lower limits of variability predicted by our RHD simulations.
This variation in our line profiles can be looked at quite naturally, since it comes directly from our models (something that cannot be done in stationary 1D models). We focus the analysis in this section on three photospheric optical lines of carbon, nitrogen, and oxygen that are often used in spectroscopic analysis of O stars.

A lower limit can be found from the mixed-sphere models (see Sect. \ref{3D_building}), since these average out the variability of individual snapshots (different assemblies of the same set of snapshots will give similar spectroscopic features).
The snapshot-sphere models do not apply such averaging, and thus provide approximate upper limits to the variability of emergent flux profiles (different assemblies of different snapshots will give different spectroscopic features). Since the snapshots used to create the snapshot-sphere models (and also mixed-sphere models) are separated in time, one could say that the line profile variability is caused by time variability of the different snapshots.

Fig. \ref{BIG_FIG}a displays results from 65 snapshot-sphere models along with the mean flux profile, illustrating a large (and very likely overestimated) level of variability in three characteristic optical absorption lines stemming from this method. Figure \ref{colored_in_O_III_snapshot_sphere} compares the typical level of variability from the two spherical reconstruction methods. 
As expected, the variability (that can also be seen in Fig. \ref{BIG_FIG}b) in the mixed-sphere models is much lower than in the snapshot-sphere models. This can be quantified by calculating the standard deviation of the variability. 
This standard deviation was calculated for each frequency-point or wavelength-point of the calculated profile. For clarity, we only provide the standard deviation (in units of normalised flux: $F_{\text{tot}}/F_{\text{cont}}$) at line-centre for each line. For the snapshot-sphere models, they are approximately 0.017 for the O III 5594 $\AA$ line, 0.018 for the C III 5697 $\AA$ line, and approximately 0.009 for the N IV 4059 $\AA$ line.  For the mixed-sphere models, these values are approximately 0.0019 for both the O III 5594 $\AA$ and the C III 5697 $\AA$ line, and 0.0009 for the N IV 4059 $\AA$ line. This indeed shows that the variability of the mixed-sphere models is much lower than the variability of the snapshot-sphere models.  
Importantly, however, the average emergent profiles agree very well between the two methods (as can be seen in Fig. \ref{BIG_FIG}c), demonstrating that for the purpose of evaluating gross 3D effects the specific averaging choice is less important.

\subsection{Averaged line profiles}

We next analysed averaged line profiles, using measurements of average full width at half maximum (FWHM) and the associated standard deviation $\sigma$, as well as the equivalent width (EW). We note that for the calculation of $\sigma$, we assume that we are working with an isotropic Gaussian: $\sigma$ = FWHM/$2\sqrt{2\ln(2)}$. The line averaging procedure is a simple averaging, where simulated data at each velocity point are added together and divided by the total number. The results of this averaging for the snapshot-spheres can be found in Fig. \ref{BIG_FIG}a, where the red line corresponds to the averaged line, and Fig. \ref{BIG_FIG}c, where the blue line corresponds to the averaged line. For the mixed-spheres, the averaged line can be seen in red in Fig. \ref{BIG_FIG}b and in red in Fig. \ref{BIG_FIG}c. Although there is a clearly visible difference in the variability of these models (as can also be seen in Fig. \ref{colored_in_O_III_snapshot_sphere}) the averages of the line profiles are almost identical. This shows that even though different model-construction-methodologies were used the average line profiles that emerge are very similar. 
The resulting FWHM, $\sigma$, and EW for the mixed-spheres can be found in Table \ref{Table_FWHM_EW_Mixed_Sphere} (see Appendix A for corresponding values for snapshot-spheres, including Tables \ref{Table_EW} and \ref{Table_FWHM} for further details). 

The results in Table \ref{Table_FWHM_EW_Mixed_Sphere} show that for the photospheric lines O III 5594 $\AA$, C III 5697 $\AA$ and N IV 4059 $\AA$ stemming from our 3D mixed-sphere models, FWHM of the averaged lines lie in between 120~km/s - 130~km/s, with a corresponding $\sigma$ in the range 52~km/s - 55~km/s. The small differences in FWHMs and $\sigma$'s between the lines could be due to them having slightly different formation regions (and hence also slightly different turbulent velocities in the formation regions). Note that these characteristic widths were obtained here directly from the 3D model without any inclusion of ad hoc micro- or macroturbulence broadening; and indeed, they are on the same order as typical observed photospheric O-star line profiles (see the introduction). 
   
 \begin{table}[h!]
\caption{FWHMs and EWs of the averaged line profiles of the mixed-spheres shown in Fig. \ref{BIG_FIG}d.}         
\label{Table_FWHM_EW_Mixed_Sphere}      
\centering                         
\begin{tabular}{c || c c c}      
\hline\hline                 %
Model-type & FWHM & $\sigma$ &  EW \\
& \multicolumn{3}{c}{(for averaged line)} \\  
\hline           
\multicolumn{4}{c}{O III 5594} \\
\hline
Mixed-sphere & & & \\
  3D & 129 km/s & 55 km/s & 10 km/s \\
  <1D> ; v = 0 & 18 km/s  & 7.5 km/s & 3.2 km/s \\
\hline     
\multicolumn{4}{c}{C III 5697} \\
\hline
Mixed-sphere & & & \\
  3D & 128 km/s & 54 km/s & 7.9 km/s \\
  <1D> ; v = 0 & 20 km/s  & 8.3 km/s & 3.4 km/s \\
\hline  
\multicolumn{4}{c}{N IV 4059} \\
\hline
Mixed-sphere & & & \\
  3D & 123 km/s & 52 km/s & 5.7 km/s \\
  <1D> ; v = 0 & 14 km/s  & 6.1 km/s & 1.9 km/s \\
\hline  

\end{tabular}
\tablefoot{See Table \ref{Table_EW} and \ref{Table_FWHM} for more details.}
\end{table}

\subsection{Comparison with 1D, static atmosphere}
We next compare the results of our synthetic photospheric line production to corresponding results of 1D codes by re-computing the lines from 1D radial averages of our snapshot-sphere and mixed-sphere models, and setting all velocities to zero.  
This approach thus mimics what would come out of an equivalent 1D, static model photosphere analysis. 
 For the mixed-sphere models, results are shown for the same three lines analysed above in Fig. \ref{BIG_FIG}d.
The FWHMs, $\sigma$'s, and EWs of the mixed-sphere models are given in Table \ref{Table_FWHM_EW_Mixed_Sphere}. As previously, corresponding values for the snapshot-sphere models can be found in Appendix A. Inspecting Fig. \ref{BIG_FIG}d and Table \ref{Table_FWHM_EW_Mixed_Sphere} one can see that, as expected, the line profiles for the 1D static model are much narrower (FWHM of the order of 10~km/s - 20~km/s instead of 120~km/s - 130~km/s), and also deeper. 
This is a direct illustration of the effect that the 3D velocity fields have upon photospheric absorption lines, offering a natural explanation for the need of `extra' ad hoc broadening mechanisms in 1D model atmosphere codes. Additionally, we note that also the EWs of the lines are different, with the ones computed from 3D models clearly higher (see discussions below). This difference in EWs shows that this broadening is not just a macroscopic broadening effect, but also affects the overall strengths of the lines (most often mimicked in 1D codes by adding ad hoc `microturbulence'). Using our 1D static model lines (by artificially broadening them to fit the 3D lines), we can find what micro- and macroturbulence velocities are needed to reproduce the 3D results. 

\begin{figure*}
   \centering
   \includegraphics[width=0.99\hsize]{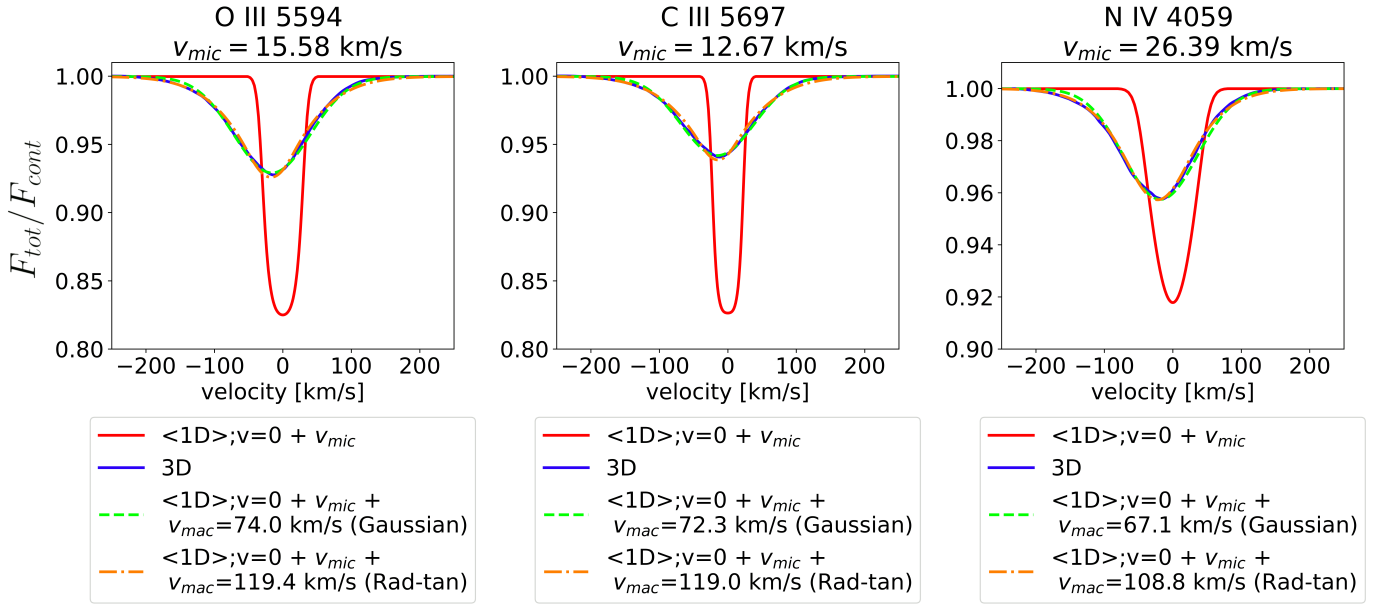}
      \caption{Micro- and macroturbulent velocities found for one mixed-sphere model, using the 1D static lines as input and the 3D lines as pseudo-observations.
              }
         \label{vmacro_fit}
   \end{figure*}

For the macroturbulence fit, we use two options: an isotropic Gaussian profile method and the radial-tangential (rad-tan) model by \cite{Gray_1975} with equal radial and tangential contributions. Fig. \ref{vmacro_fit} shows that for isotropic macroturbulence we find best-fit values $\varv_{\rm vmac} \sim$\,70~km/s. This is consistent with the discussion above regarding velocity dispersion, since the characteristic velocity in this macroturbulence model is directly related to $\sigma$ by $\varv_{\rm mac} = \sigma \sqrt{2}$. For the rad-tan model we as expected find higher characteristic velocities ($\sim$\,110-120~km/s); it is a well-known result (e.g. \citealt{Gray_1975, Sundqvist_2014}) that an isotropic macroturbulence model returns lower characteristic velocities than the rad-tan method. To reproduce the EWs of the 3D lines we need to apply an isotropic microturbulence $\sim$\,10-20~km/s, where the different best-fit numbers for the different lines could reflect different formation heights or be a saturation effect caused by our neglect of scattering in the calculation of the source function (see discussion below). We note further that it is currently unclear to what extent 3D effects on EWs can be fully captured by such microturbulence, or whether it rather is indicative of 3D chemical abundance effects (see discussion in Section \ref{Disc_and_conc}).
Another important note is that the lines in Fig. \ref{BIG_FIG}d are slightly shifted towards the red in comparison to the 1D averaged lines. The shift values were approximately -17~km/s for the O III 5594 $\AA$ line, -15~km/s for the C III 5697 $\AA$ line, and -20~km/s for the N IV 4059 $\AA$ line. This shift is due to the fact that the photospheric surface is a dynamic one, and as such, the emitting surface is shifting the line. This net-infall of material around the photosphere was noted also in \citet{Dwaipayan} and can be seen in the middle panel of Fig. \ref{3d_slice}, which shows that more than half of the surface has infalling material.

\section{First line profile results including scattering} \label{First_line_profile_results_including_scattering}
This section discusses first test results when line profiles are computed using scattering-dominated source functions.
Basic assumptions and methods of this are the following (see \citealt{Levin_2020}). 
The continuum and line source functions $S_C$ and $S_L$ are, respectively:

\begin{equation}
    S_C = (1-\epsilon_C)J_\nu + \epsilon_CB_\nu,
\end{equation}
\vspace{-15pt}
\begin{equation}
    S_L = (1-\epsilon_L)\bar{J} + \epsilon_L B_{\nu_0},
\end{equation}

with $\epsilon_C$ and $\epsilon_L$ the thermalisation parameters (0 is full scattering, 1 is full thermal), $J_\nu$ the mean intensity, $\bar{J}$ the profile-weighted mean line intensity and $B_{\nu_0}$ the Planck function at line-centre $\nu_0$.
The determination of source functions assumes a two-level atom with a thermal+scattering continuum for the line formation.
For these first tests, we simply took both $\epsilon$ values as fixed input parameters (work is underway to instead compute them from atomic line and continuum data).  
Source functions are derived from  3D short-characteristics (SC) solutions to the equations of radiative transfer, discretised on a non-uniform Cartesian grid. The solution is augmented by an ‘accelerated-lambda-iteration’ (ALI) scheme, using non-local operators to ensure convergence. The procedure accounts fully for Doppler shifts from arbitrary 3D non-relativistic velocity fields, as described in detail by \citet{Levin_2020}. The line opacities were still calculated using the approximate NLTE approach described above.
The Cartesian grid used for these calculations consisted of 219 points along each axis (so 110 in each positive or negative direction of x, y, and z). Using this part of our code package further allows us to compare short-characteristics (SC) fluxes to corresponding fluxes predicted by the analytic FLD relation used in our RHD simulations. In this respect, however, it is important to keep in mind that the strong frequency-dependent nature of fluxes makes it computationally prohibitive to apply the SC technique directly in the RHD simulations, at least in cases where opacity from a multitude of spectral lines is important. As such, we here instead compared FLD and SC fluxes by applying a simplified gray opacity model, and present results from this in Appendix B (see also Section 5.2. in \citealt{nico_2022b} for more such comparisons).

\begin{figure}[]
   \centering
   \includegraphics[width=0.99\hsize]{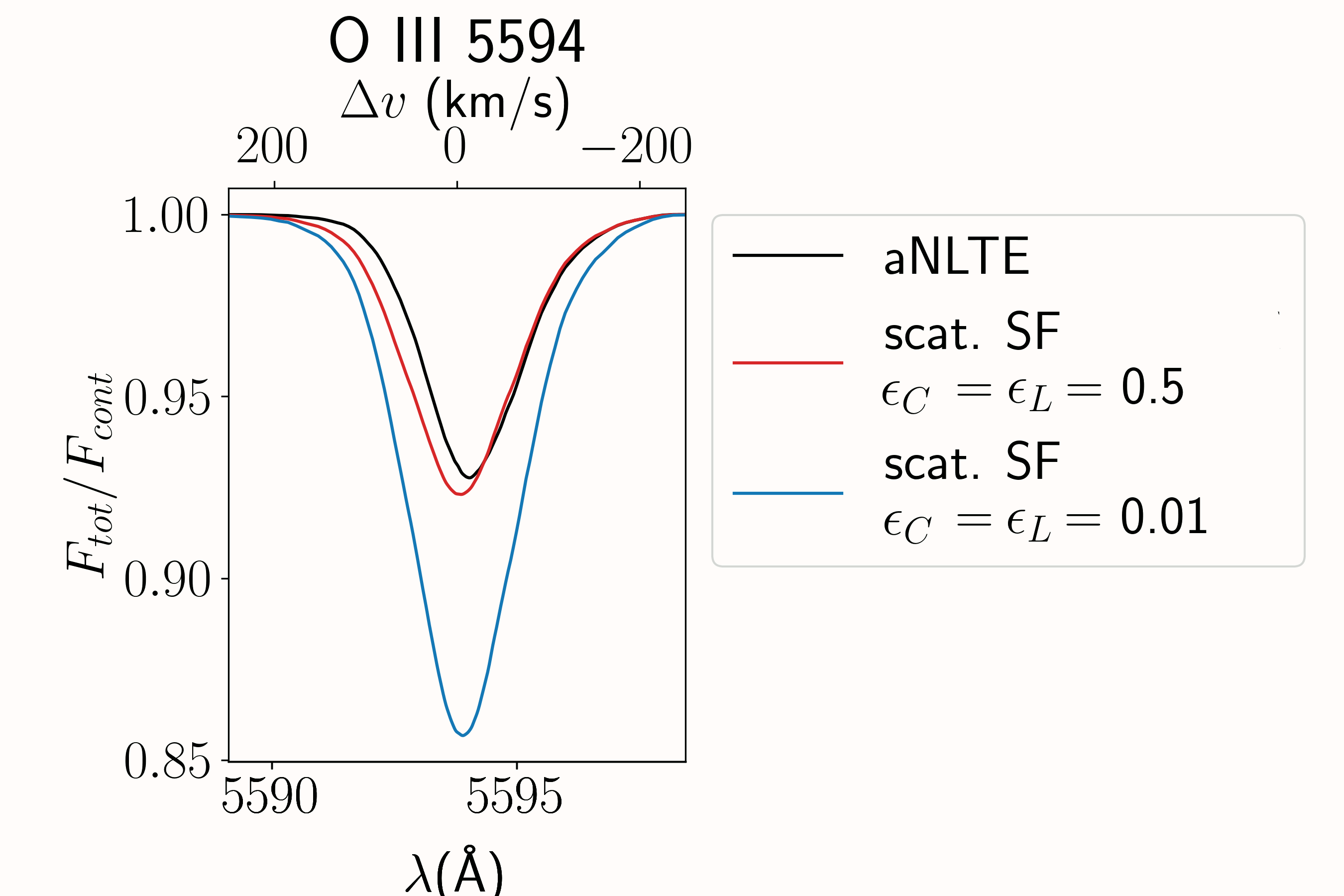}
      \caption{O III 5594 $\AA$ line for one mixed-sphere model, using two different calculation methods. First, the previously shown aNLTE approach (the black line). Second, the scattering-dominated source function approach (red and blue line). Note that the difference between the red and blue line is the used values of the thermalisation parameters $\epsilon_C$ and $\epsilon_L$. 
              }
         \label{O_III_scat_source_function}
   \end{figure}

Including scattering typically reduces source functions in the line forming layers. As a consequence, this could potentially desaturate photospheric lines that may be saturated in our previous results. This is important for future applications, for example comparing synthetic lines to observations, since it can affect the line depths (and hence EWs). We note, however, that in our tests the FWHMs and $\sigma$ derived above were barely affected by the inclusion of scattering. 
The deepening effect on the line stemming from scattering can be seen in Fig. \ref{O_III_scat_source_function} for the O III 5594 $\AA$ line, which shows that the scattering line source function is significantly depressed as compared to the aNLTE source function.

\begin{figure}[]
   \centering
   \includegraphics[width=0.99\hsize]{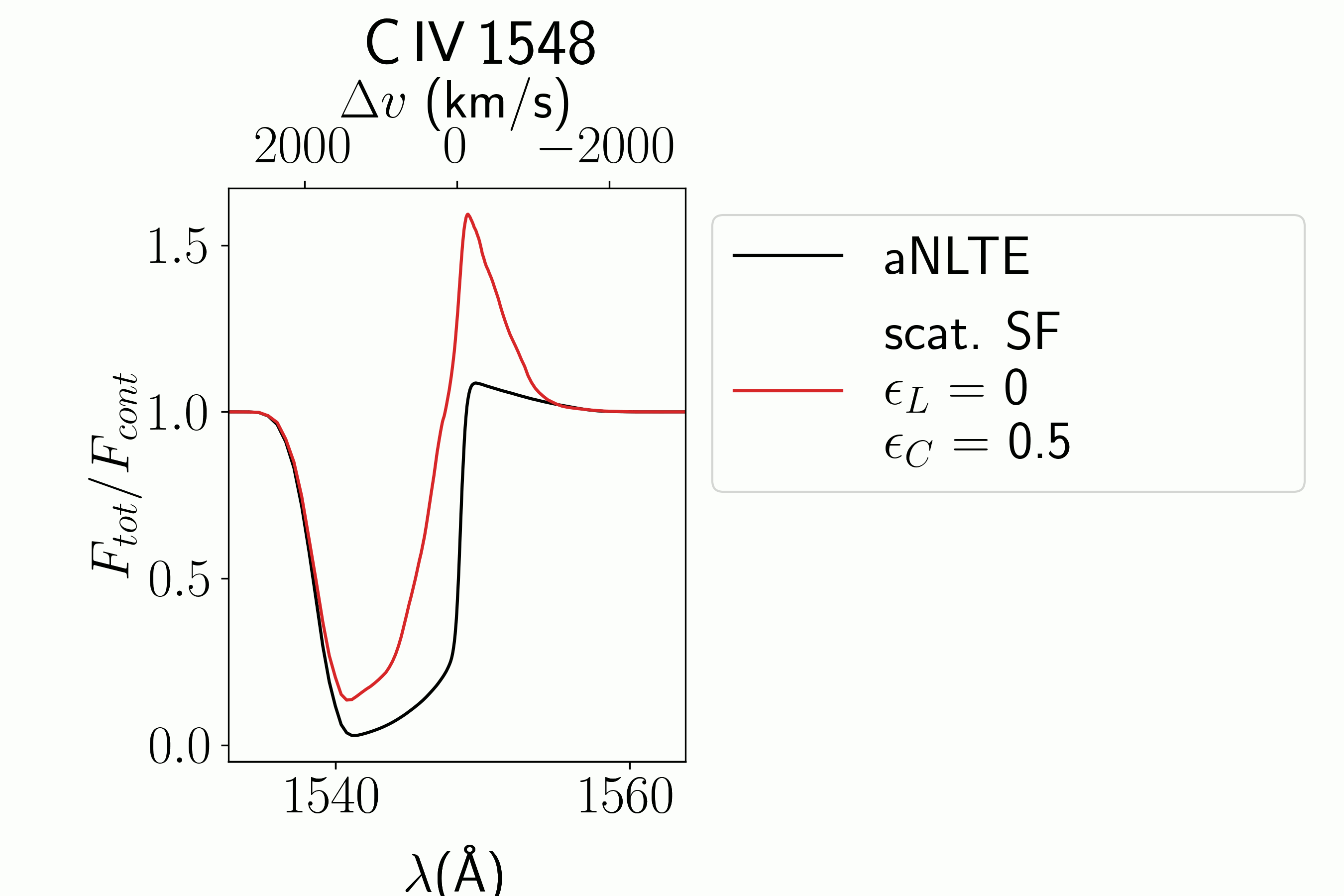}
      \caption{C IV 1548 $\AA$ line for one mixed-sphere model, using two different calculation methods. First, the previously shown aNLTE approach (the black line). Second, the scattering-dominated source function approach (red line).
              }
         \label{C_IV_1s2_2s_scat_source_function}
   \end{figure}

The scattering line source function also allows us to model the strong UV resonance lines that typically appear as P-Cygni lines in O-star spectra. Figure \ref{C_IV_1s2_2s_scat_source_function} shows the carbon UV resonance line at 1548 $\AA$. 
In this figure, it is clearly visible that we have a much better P-Cygni wind profile compared to our aNLTE method.  Note, however, that the underlying 3D RHD O-star simulation did not go out far enough in radius to capture the complete C IV 1548 $\AA$ line formation region, which means that we may expect the blue edge of the line to reach somewhat higher velocities than in this figure. Moreover, since the investigated C IV line in reality is a doublet, this must be accounted for before a direct comparison to observations can be done. Nonetheless, the figure illustrates how our 3D models are able to naturally capture key elements of such P-Cygni lines, such as the softening of the blue absorption edge by the turbulent velocity field.
In summary, Figs. \ref{O_III_scat_source_function} and \ref{C_IV_1s2_2s_scat_source_function} show first promising results from using scattering-dominated source functions, taking us one step closer to performing quantitative spectroscopy using 3D models.

\section{Discussion and conclusion} \label{Disc_and_conc}

When inspecting our 3D RHD simulation of a typical early O star in the Galaxy, we found turbulent velocities of approximately 48~km/s, alternatively measured as velocity dispersions ($\sigma_{vd}^2=<v_r^2> - <v_r>^2$) 
of approximately 57~km/s at the average optical photosphere.
Computing averaged photospheric absorption line profiles (see Table \ref{Table_FWHM_EW_Mixed_Sphere}), we found broadening with standard deviation of approximately 54~km/s. That is, we may directly relate the magnitudes of turbulent velocities measured in the RHD simulations to the typical broadening expected for corresponding absorption line profiles. Moreover, as our fits to corresponding 1D models applying microturbulence and macroturbulence show, the characteristic photospheric velocities in our 3D models are broadly on the same order as typically observed for O stars (for example \citealt{Sim_n_D_az_2010, Simon-Diaz2014, Simon-Diaz2017, Nadya2023, Nadya2024}).

This large natural broadening of massive-star absorption lines has been found in other recent 3D modelling as well (\citealt{Schultz_2023}, see for example their Figs. 9 and 10), using Monte-Carlo radiative transport to produce absorption lines from the 3D ATHENA++ RHD simulations by \citet{Jiang_2015, Schultz_2022}. Since their chosen massive stars to model did not have the same fundamental parameters as the O star presented here, the broadening results can however not be directly quantitatively compared. Also, in addition to different radiative transfer techniques, the underlying 3D RHD simulation methods differ between the two groups, as their ATHENA++ framework on the one hand uses a more sophisticated radiation closure (based on an Eddington tensor method), but on the other hand does not account for line driving effects (thus their simulations lack a wind outflow). Nonetheless, the overall qualitative agreement in results stemming from these two independent developments is very reassuring for future 3D massive-star model atmosphere work. 

In addition to line broadening, the average atmospheric properties also change in multi-D simulations. Perhaps most strikingly for O stars, the strong turbulent pressure results in much larger photospheric scale-heights, and thus in significantly shallower density profiles (see also \citealt{Dwaipayan}).   
As recently investigated by \citet{Gemma2025} this affects in particular surface gravity determinations from observed spectral line profiles,   
and it is suggested that this might present a solution to the so-called `mass-discrepancy problem' that is seen between evolutionary and spectroscopy mass determinations (\citealt{Herrero_1992}).
In this paper we have demonstrated how the photospheric turbulent velocity is directly related to the measured dispersion of absorption lines. While awaiting full 3D-based spectroscopic studies, one can thus use observed velocity dispersions in photospheric lines (at least in the absence of dominant rotational broadening) to calibrate the turbulent velocity needed to correct the hydrostatic equations used in all present-day 1D atmospheric models.

Finally, we note that we have potential 3D effects on abundance determinations, since the EWs found for the lines are different than in corresponding 1D models. This may lead to differences in derived abundances for the same objects, although the relative role of such 3D abundance corrections and the `microturbulence' used in 1D models is not clear at the present (also since this strongly depends on the treatment of line scattering which will be further studied in future work).  
In comparison to studies on low-mass stars, abundance determinations done using 3D (time-dependent hydrodynamical) models of the Sun's atmosphere have even changed the solar chemical composition yardstick \citep{Asplund_2009}. 
For such low-mass stars in general, studies over the past decade have shown that results derived from 1D average structures are not adequate as substitutes for full 3D spectroscopy (\citealt{Lind_2024}). 

In this paper we have shown first promising results towards spectroscopy using 3D models also for hot, massive stars. In future work, we shall extend our formalism to include doublets, damping wings, as well as full scattering and NLTE effects, enabling a direct comparison to observations by means of quantitative spectroscopy. 

\begin{acknowledgements}

The computational resources used for this work were provided by Vlaams Supercomputer Centrum (VSC) funded by the Research Foundation-Flanders (FWO) and the Flemish Government. The authors gratefully acknowledge support from the European Research Council (ERC) Horizon Europe under grant agreement number 101044048, from the Belgian Research Foundation Flanders (FWO) Odysseus program under grant number G0H9218N, from FWO grant G077822N, and from KU Leuven C1 grant BRAVE C16/23/009. The authors would also like to thank previous members of the KUL-EQUATION group for their earlier contributions. Finally, the authors would like to thank all members of the KUL-EQUATION group for fruitful discussion, comments, and suggestions. LD would also like to thank Conny Aerts for giving feedback on the draft, Levin Hennicker for the help and discussions in order to understand the general code package better and Stan Owocki for the discussions and insights. We thank the referee for their useful comments on our manuscript. We made significant use of the following packages to analyse our data: {\fontfamily{qcr}\selectfont NumPy} \citep{harris_2020}, {\fontfamily{qcr}\selectfont SciPy} \citep{virtanen_2020}, {\fontfamily{qcr}\selectfont matplotlib} \citep{hunter_2007}, {\fontfamily{qcr}\selectfont Python amrvac\_reader} \citep{keppens_2020}, {\fontfamily{qcr}\selectfont PyVista} \citep{pyvista}.

\end{acknowledgements}
\newpage

%
%

\bibliographystyle{aa}
\bibliography{references}

\begin{appendix}
    
\setcounter{table}{0}
\renewcommand{\thetable}{A\arabic{table}}
\section*{Appendix A: Additional tables}

\begin{table}[h]
\caption{Summary of EWs of the line profiles computed in this work.}             
\label{Table_EW}      
\centering                         
\begin{tabular}{c || p{24mm} p{28mm} p{24mm}  }      
\hline\hline                 %
Model-type & Average EW \newline (individual lines) & [Min; Max] EW \newline (individual lines) & EW \newline (averaged line) \\   
\hline           
\multicolumn{4}{c}{} \\
\multicolumn{4}{c}{O III 5594} \\
\multicolumn{4}{c}{} \\
\hline
Snapshot-sphere & & &  \\
&  & &   \\ 
  3D & 10.03 km/s & [7.15; 12.49] km/s \newline (range = 5.34 km/s) & 9.97 km/s  \\
  <1D> ; v = 0 & 3.48 km/s & [2.45; 4.34] km/s \newline (range = 1.89 km/s) & 3.48 km/s \\
  &  & &   \\
  Mixed-sphere & & &  \\
  &  & &   \\ 
 3D & 9.91 km/s & [9.53; 10.09] km/s \newline (range = 0.56 km/s) & 9.9 km/s   \\
  <1D> ; v = 0 & 3.19 km/s & [3.04; 3.3] km/s \newline (range = 0.26 km/s) &  3.19 km/s \\
\hline     
\multicolumn{4}{c}{} \\
\multicolumn{4}{c}{C III 5697} \\
\multicolumn{4}{c}{} \\
\hline
Snapshot-sphere & & &  \\
&  & &   \\ 
  3D & 8.27 km/s & [4.52; 10.83] km/s \newline (range = 6.31 km/s) & 8.23 km/s \\
  <1D> ; v = 0 & 3.79 km/s & [2.63; 4.85] km/s \newline (range = 2.22 km/s) & 3.79 km/s \\
  &  & &   \\
  Mixed-sphere & & &  \\
  &  & &   \\ 
 3D & 7.9 km/s & [7.5; 8.19] km/s \newline (range = 0.69 km/s) & 7.89 km/s \\
  <1D> ; v = 0 & 3.42 km/s & [3.26; 3.56] km/s \newline (range = 0.3 km/s) & 3.42 km/s \\
 \hline     
\multicolumn{4}{c}{} \\
\multicolumn{4}{c}{N IV 4059} \\
\multicolumn{4}{c}{} \\
\hline
Snapshot-sphere & & &  \\
&  & &   \\ 
  3D & 5.66 km/s & [4.38; 7.6] km/s \newline (range = 3.22 km/s) & 5.62 km/s \\
  <1D> ; v = 0 & 2.06 km/s & [1.11; 2.77] km/s \newline (range = 1.66 km/s) & 2.06 km/s \\
  &  & &   \\
  Mixed-sphere & & &  \\
  &  & &   \\ 
 3D & 5.69 km/s & [5.59; 5.77] km/s \newline (range = 0.18 km/s) & 5.69 km/s \\
  <1D> ; v = 0 & 1.86 km/s & [1.8; 1.93] km/s \newline (range = 0.13 km/s) & 1.86 km/s \\
 \hline     
\end{tabular}
\end{table}

\begin{table*}[h]
\caption{Summary of FWHMs of the line profiles computed in this work.}             
\label{Table_FWHM}      
\centering                         
\begin{tabular}{c || p{24mm} p{31mm} p{24mm}|p{21mm} p{21mm}  }      
\hline\hline                 %
Model-type & Average FWHM \newline (individual lines) & [Min; Max] FWHM \newline (individual lines) & Average $\sigma$ \newline (individual lines) & FWHM \newline (averaged line) & $\sigma$ \newline (averaged line)  \\   
\hline           
\multicolumn{6}{c}{} \\
\multicolumn{6}{c}{O III 5594} \\
\multicolumn{6}{c}{} \\
\hline
Snapshot-sphere & & & & & \\
&  & &  &  & \\ 
  3D & 123.88 km/s  & [81.71; 182.84] km/s \newline (range = 101.12 km/s) & 52.61 km/s & 132.72 km/s & 56.36 km/s \\
  <1D> ; v = 0 &  17.64 km/s  & [15.87; 18.53] km/s \newline (range = 2.66 km/s) &  7.49 km/s & 17.6 km/s & 7.47 km/s \\
  &  & &  &  & \\ 
  Mixed-sphere & & & & & \\
  &  & &  &  & \\ 
  3D & 128.41 km/s  & [118.8; 138.31] km/s \newline (range = 19.51 km/s) & 54.53 km/s & 129.26 km/s  & 54.89 km/s \\
  <1D> ; v = 0 & 17.6 km/s  & [17.6; 17.6] km/s \newline (range = 0.0 km/s) &  7.47 km/s & 17.6 km/s &  7.47 km/s \\
\hline     
\multicolumn{6}{c}{} \\
\multicolumn{6}{c}{C III 5697} \\
\multicolumn{6}{c}{} \\
\hline
Snapshot-sphere & & & & & \\
&  & &  &  & \\ 
3D & 122.12 km/s  & [82.42; 183.06] km/s \newline (range = 100.64 km/s) & 51.86 km/s & 130.99 km/s & 55.63 km/s \\
<1D> ; v = 0 &  19.5 km/s  & [16.71; 21.6] km/s \newline (range = 4.89 km/s) & 8.28 km/s & 19.5 km/s & 8.28 km/s \\
   Mixed-sphere & & & & & \\
  &  & &  &  & \\ 
 3D & 126.3 km/s  & [115.11; 136.66] km/s \newline (range = 21.55 km/s)  & 53.63 km/s & 127.57 km/s  & 54.17 km/s \\
 <1D> ; v = 0 &  19.5 km/s  & [19.5; 19.5] km/s \newline (range = 0.0 km/s) & 8.28 km/s & 19.5 km/s & 8.28 km/s \\
 \hline     
\multicolumn{6}{c}{} \\
\multicolumn{6}{c}{N IV 4059} \\
\multicolumn{6}{c}{} \\
\hline
Snapshot-sphere & & & & & \\
&  & &  &  & \\ 
3D & 118.7 km/s  & [71.17; 178.9] km/s \newline (range = 107.73 km/s) & 50.41 km/s & 128.18 km/s & 54.43 km/s \\
 <1D> ; v = 0 & 14.75 km/s  & [13.57; 15.87] km/s \newline (range = 2.3 km/s) & 6.26 km/s & 15.07 km/s & 6.4 km/s \\
  &  & &  &  & \\ 
  Mixed-sphere & & & & & \\
  &  & &  &  & \\ 
  3D & 125.47 km/s  & [120.66; 137.92] km/s \newline (range = 21.55 km/s)  & 53.28 km/s & 122.77 km/s  & 52.14 km/s \\
   <1D> ; v = 0 & 14.3 km/s  & [14.3; 14.3] km/s \newline (range = 0.0 km/s) & 6.07 km/s & 14.3 km/s & 6.07 km/s \\
 \hline     
\end{tabular}
\end{table*}

\newpage
\clearpage
\section*{Appendix B: Short-Characteristics method versus Flux Limited Diffusion method}

\setcounter{figure}{0}
\renewcommand{\thefigure}{B\arabic{figure}}

Here, we compare the Short-Characterics (SC) method that our radiative transfer code package uses for calculation of the source function in section \ref{First_line_profile_results_including_scattering} with the Flux Limited Diffusion (FLD) method that our RHD model atmospheres with winds simulations use.  
This is done in the following way: 
we take one snapshot, and apply our SC method on this snapshot to find the frequency-integrated Eddington flux using a pure absorption gray Thomson opacity model and the same 3D Cartesian technique as the corresponding SC calculations presented in \citet{nico_2022b}. From this snapshot data, we then also derive the Eddington flux using the analytic FLD approximation as outlined in \citet{nico_2022a}. 

\begin{figure}[h!]
   \centering
   \includegraphics[width=0.99\hsize]{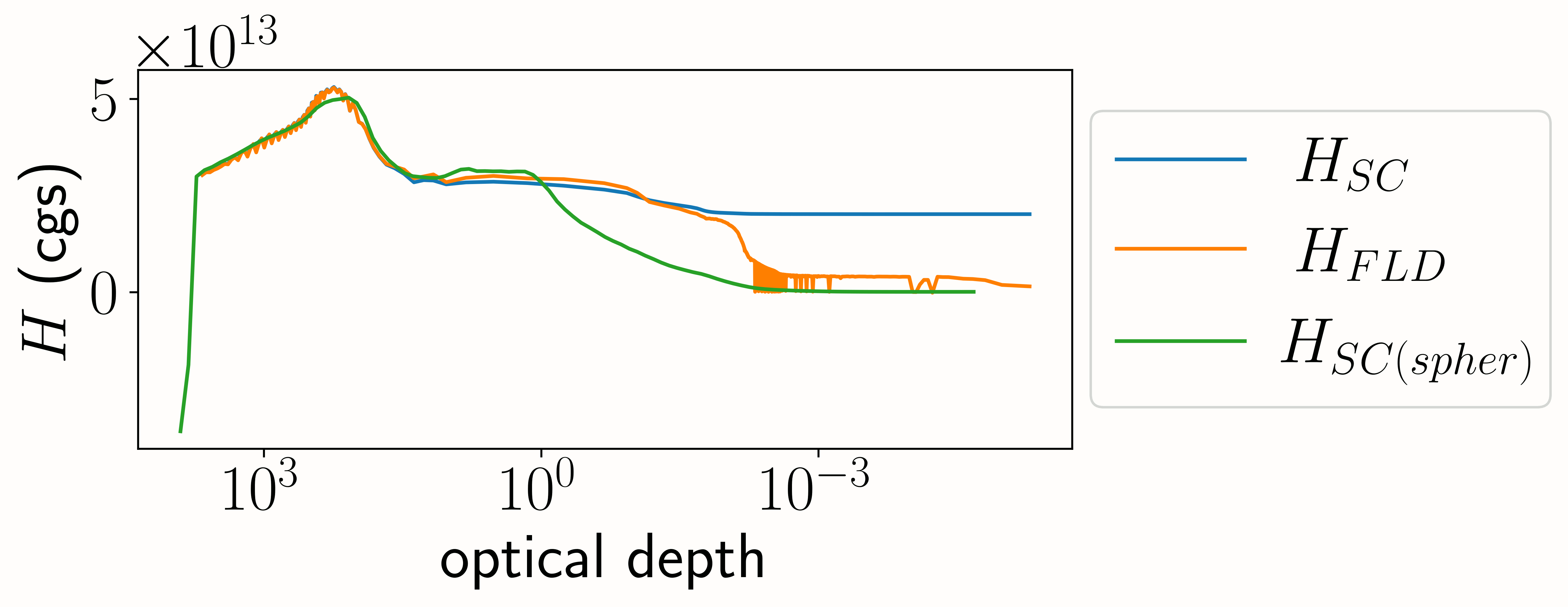}
      \caption{1D radial averages of the Eddington flux, using an SC method, an FLD method, and a translated-from-spherical-SC method. The optical depth on the x axis is the average continuum optical depth.}
         \label{1D_radial_av_snapshots}
   \end{figure}

\begin{figure*} 
       \centering
\includegraphics[width=0.9\linewidth]{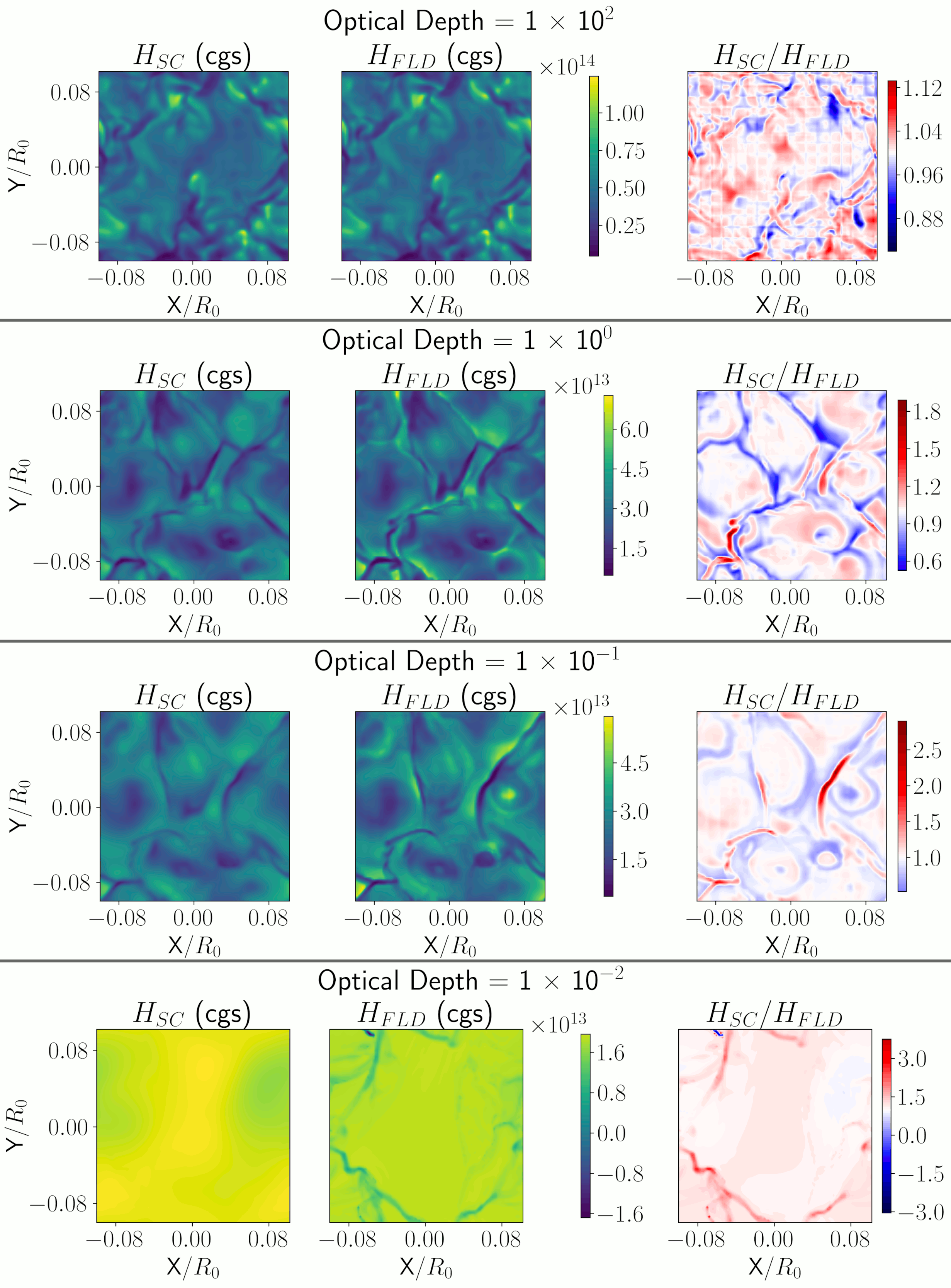}
      \caption{The Eddington flux calculated using the SC method, FLD method, and the ratio of them, for different vertical planes in the 3D RHD model with average continuum optical depths according to the labels. The lateral X and Y axes are labelled in units of $R_0$.}
         \label{tau_slices_snapshot} 
   \end{figure*}

We also compute Eddington fluxes using a third method, namely SC solutions applied to the corresponding snapshot-sphere model taking full sphericity effects into account. However, these Eddington fluxes cannot directly be compared to the RHD results, since this part of our radiative transfer code package (that deals with complete spherical models) is developed for post-processing purposes and so only contains frequency-dependent quantities.
To remedy this, we set a $T_{rad,\nu}$ using the Planck function 
from the SC Eddington fluxes of the corresponding snapshot-sphere model, calculate a frequency-integrated mean intensity $J = \frac{\sigma}{\pi}T_{rad}^4$ from this, and use this mean intensity to determine our Eddington flux using the analytic FLD approximation. This gives a rough indication of full sphericity effects upon the fluxes. The results of these comparisons can be seen in Fig. \ref{1D_radial_av_snapshots}, where we have taken radial averages. 
Fig. \ref{1D_radial_av_snapshots} shows that in the inner layers of the atmosphere we find nice agreement between the different methods. However, for the outer optically thin wind layers there are clear differences. When inspecting the $H_{FLD}$ curve we can see that the spherical correction we apply in the RHD simulations is, on average, ‘intermediate’ between the Cartesian SC result ($H_{SC}$)
and the full spherical SC result ($H_{SC(spher)}$, which clearly displays the expected $1/r^2$ spherical dilution effect). Of course, in addition to sphericity one must not forget that the FLD closure relation itself will have a direct effect on the difference between SC and FLD fluxes; this will require further work in order to improve our closure relation in the RHD simulations.
In Fig. \ref{tau_slices_snapshot} we show slices of the 3D RHD model at different optical depths for a given snapshot, focusing now on direct comparison to the 3D Cartesian SC fluxes. In deeper layers, there are
only relatively small differences present, and in particular the main structures of the FLD and SC fluxes are very similar. In the outer optically thin layers, on the other hand, the SC flux becomes larger than the FLD flux; this again arises due to a combination of the approximate FLD closure and sphericity effects. Moreover, the SC fluxes in these outer regions are significantly smoother than the FLD fluxes; this is because the SC method integrates over large distances whereas the FLD flux is based on a local gradient approach.

\end{appendix}

\end{document}